\documentclass[aps,twocolumn,showpacs,floats,prb,floatfix]{revtex4}
\usepackage{epsf}
\usepackage{float}
\usepackage{amsmath}
\usepackage{graphicx}

\newcommand{\vp}{{\bf p}}
\newcommand{\me}{\phantom{a}_{\textrm{out}}\langle \vp\;\sigma|\vp'\;\sigma'\rangle_{\textrm{in}}}

\newcommand{\vx}{{\bf x}}
\newcommand{\vy}{{\bf y}}

\newcommand{\ves}{,}

\newcommand{\be}{\begin{equation}}
\newcommand{\ee}{\end{equation}}
\newcommand{\bea}{\begin{eqnarray}}
\newcommand{\eea}{\end{eqnarray}}

\newcommand{\laci}{}

\begin{document}

\title {
Theory of inelastic scattering from quantum impurities
}

\author{L\'aszl\'o Borda$^1$, Lars Fritz$^{2,3}$, Natan Andrei$^4$, and  Gergely Zar\'and$^{5}$}
\affiliation{
$^1$ 
Research Group ``Theory of Condensed Matter'' of the  Hungarian Academy of Sciences
\\
 Budapest University of Technology and Economics\ves
Budafoki \'ut 8.  H-1521 Hungary\\
$^2$ Institut f\"ur Theorie der Kondensierten Materie and DFG-Center for
Functional 
Nanostructures (CFN)\ves
Universit\"at Karlsruhe\ves 76128 Karlsruhe\ves Germany \\
$^3$Institut f\"ur Theoretische Physik,
Universit\"at zu K\"oln, Z\"ulpicher Str. 77, 50937 K\"oln, Germany\\
$^4$ Center for Materials Theory, Rutgers University, Piscataway, NJ 08855, U.S.A. \\
$^5$   Department of Theoretical Physics\ves Budapest University of Technology and Economics\ves
Budafoki \'ut 8.  H-1521 Hungary}
\date{\today}

\begin{abstract}
We use the framework set up recently to compute non-perturbatively  
inelastic scattering from quantum impurities [G. Zar\'and {\it et al.},
Phys. Rev. Lett. {\bf 93}, 107204 (2004)] to 
study
the the energy dependence of the single 
particle $S$-matrix and the inelastic scattering cross section  for a number of quantum impurity 
models. We study the case of  the spin $S=1/2$  two-channel Kondo model, the
Anderson model, and the usual $S=1/2$ single-channel Kondo model.  
We discuss the difference
between non-Fermi liquid and Fermi liquid models and study  how a 
cross-over between the non-Fermi liquid and Fermi liquid regimes appears in
case of channel anisotropy for the $S=1/2$ two-channel Kondo model. 
We show that for the most elementary non-Fermi liquid system, the two-channel
Kondo model, half of the scattering
remains inelastic even at the Fermi energy. 
Details of the derivation of the reduction formulas 
and a simple path integral approach to connect the $T$-matrix to local
correlation functions are also presented.
\end{abstract}


\maketitle


\section{Introduction}

Quantum interference effects play a major role in mesoscopic systems: they 
lead to phenomena such as Aharonov-Bohm interference~\cite{AB_int}, weak
localization effects~\cite{Birge2003,Saminadayar2003}, 
universal conductance fluctuations~\cite{UCF}, or mesoscopic local 
density of states fluctuations~\cite{Altshuler_review}.
All these interesting phenomena rely on the phase coherence of the conduction electrons. 
This phase coherence is,  however, 
destroyed through {\em inelastic scattering processes}, where some 
excitation
is created in the {\em environment}, and which thus lead to a loss of quantum interference 
after a characteristic time, $\tau_\varphi$. This characteristic time is called
the dephasing time or sometimes as the inelastic scattering time. 
The excitations created in course of an inelastic scattering process may be phonons, magnons, 
electromagnetic  radiation, or simply electron-hole excitations, where in the latter case 
the 'environment' is provided by   the conduction electrons themselves. 

A few years ago Mohanty and Webb measured  the dephasing time
$\tau_\varphi(T)$ very carefully down to very low temperatures  through weak localization
experiments, and reported a surprising saturation of it at the lowest temperatures.~\cite{Mohanty-Webb,mohanty2003}  
These experiments gave rise to many theoretical speculations:  
 intrinsic dephasing due to electron-electron interaction,~\cite{Zaikin,Aleiner,Delft}  
scattering from two-level systems,~\cite{Zawa,Imry} 
and even coupling to zero point fluctuations
have been proposed to explain 
the observed saturation, and induced rather harsh discussions.
{\laci
Finally, it has been  shown recently
that an apparent saturation 
can also appear due to
inelastic scattering 
from 
magnetic impurities.~\cite{Birge2003,zar_inel} 
}

Triggered  by these results of Mohanty and Webb, a number of experimental groups
also revisited the problem of inelastic scattering and dephasing in quantum wires
and disordered metals: A series of experiments have been performed
to study   the non-equilibrium relaxation 
of the energy distribution function in short 
quantum wires of various compositions.~\cite{Saclay}
Depending on the material, these energy relaxation experiments could be well explained in terms of  
the orthodox theory of electron-electron  interaction in one-dimensional wires,~\cite{AAK} 
and/or inelastic scattering mediated by magnetic
impurities.~\cite{Abrikosov,Zawadowski,Glazman,Aleiner,Altshuler,Kroha}
Parallel to, and partially triggered by these experiments, a systematic study
of the inelastic scattering from 
magnetic impurities has also been carried out recently, where inelastic scattering 
at energies down to well below the Kondo scale has also been studied.~\cite{Saminadayar2005,Saminadayar2006,Birge2006} 
Describing inelastic scattering from magnetic impurities around and below the Kondo scale
has been a theoretical challenge, since this regime can be reached only 
through  {\em non-perturbative} methods.  This goal has been finally reached 
in Refs.~\onlinecite{zar_inel} and \onlinecite{Rosch}: 
In Ref.~\onlinecite{zar_inel}  a theory of inelastic scattering  has been developed
at $T=0$ temperature, while the authors of Ref.~\onlinecite{Rosch} showed  
that the finite temperature version of the formula introduced in 
Ref.~\onlinecite{zar_inel} describes the dephasing rate that appears in 
the expression of weak localization in the limit of small concentrations too. 
Except for very low temperatures,  where a small residual 
inelastic scattering is observed,~\cite{Saminadayar2006,Birge2006} these calculations
are in very good agreement with the experiments 
that clearly show that magnetic impurities in 
concentration 
as small
as 1ppm 
can induce  substantial inelastic scattering.~\cite{zar_inel,Rosch,Achim2}   
{\laci
The source of the small residual inelastic scattering is unclear:
It may be due to some mispositioned magnetic impurities with anomalously
small Kondo temperature or structural defects created by the ion implantation,
but an intrinsic effect cannot be excluded either, although the residual
dephasing seems to be proportional to the impurity concentration.
}
Furthermore, we have to emphasize, 
that other experiments on very dirty metals probably cannot be explained 
in terms of magnetic scattering, 
and 
possibly other mechanisms are needed to
account for the dephasing observed at very low temperatures
in these systems.~\cite{Lin}

The purpose of the present paper is to demonstrate, how the rather general 
theory of Ref.~\onlinecite{zar_inel} can be applied to various quantum
impurity problems. 
In our previous  work we presented results only for  the single channel 
$S=1/2$
Kondo  model, while  in the  present paper  we extend  our study  to different
quantum impurity  models (two  channel 
$S=1/2$ Kondo  model, Anderson  model) as
well, and we  also discuss some analytical expressions  for various scattering
rates in the single channel Kondo  model. In addition, we present many details
of derivation of the formalism shortly discussed in Ref.~\onlinecite{zar_inel}.

 Ref.~\onlinecite{zar_inel} formulates the problem of inelastic scattering 
in terms of the  many-body $S$-matrix defined through the overlap of incoming and outgoing
scattering states:
\begin{eqnarray}
\phantom{a}_\textrm{out}\langle f |i\rangle_\textrm{in}
\equiv \phantom{a}_\textrm{in}\langle f
|\hat S|i\rangle_\textrm{in}\;.
 \end{eqnarray}
The incoming and outgoing scattering states, $|i\rangle_\textrm{in}$ and
$|f\rangle_\textrm{out}$,  are asymptotically free, and they may contain many excitations, 
i.e. they are true many-body states. 
In the interaction representation $\hat S$ is given by the well-known expression
\begin{eqnarray}
\hat S=T \textrm{exp}\left [ -i \int_{-\infty}^{\infty} H_{\textrm{int}}(t)\textrm{d}t\right],
 \end{eqnarray}
 with $T$ the usual time-ordering operator, and $H_{\textrm{int}}$ the
 interaction part that induces scattering. 

The many-body $T$-matrix is defined as the
 'scattering part' of the $S$-matrix,
 \begin{eqnarray}
 \hat{S}=\hat{I}+i\hat{T},
 \end{eqnarray}
 where $\hat{I}$ denotes the identity operator.
Energy conservation implies that
\be
_\textrm{in}\langle f |\hat T|i\rangle_\textrm{in}  = 2\pi \; \delta(E_f-E_i)
\; \langle f | {\cal T}|i\rangle \; ,
\label{eq:on_shellT}
\ee
with the $\langle f |{\cal T}|i\rangle $ the on-shell $T$-matrix.

\begin{figure}[htb]
\includegraphics[width=0.9\columnwidth,clip]{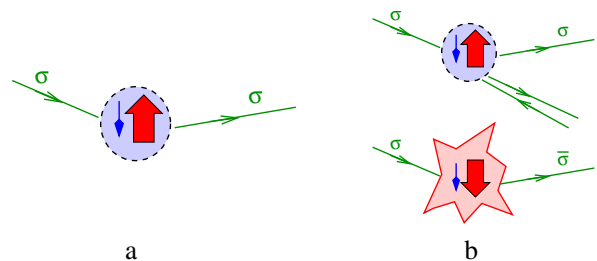}
\caption{\label{fig:ineastic_elastic}
{Sketch of (a) elastic and (b) inelastic scattering processes.
In case of an inelastic scattering the outgoing 
electron leaves spin- and/or electron-hole excitations behind.}
}
\end{figure}

The results of Ref.~\onlinecite{zar_inel} rely on the simple observation, that 
the on-shell matrix elements of the many-body  $T$-matrix
between single particle states,  
$\langle {\bf p}\sigma |{\cal T}| {\bf p'}\sigma' \rangle$,
determine both the total  ($\sigma_{\rm tot}$) and the 
elastic 
($\sigma_{\rm el}$)
scattering cross sections of the conduction
electrons (or holes) at $T=0$ temperature. 
The total scattering cross section of an electron of momentum 
${\bf p}$ and spin $\sigma$ is given by the optical theorem as
 \be
\sigma_{\rm total}^\sigma = \frac{2}{v_F} {\rm Im} \langle {\bf p} \sigma | {\cal T} | {\bf p} \sigma\rangle
\;, 
\label{eq:sigma_tot}
\ee
where the velocity of the incoming electron is approximated by the Fermi
velocity, $v_F$. 
In case of elastic scattering, an incoming single electron state is
scattered into an outgoing single electron state, without inducing any 
spin or electron-hole excitation  of the environment. The 
corresponding cross section can be expressed as
\be
\sigma_{\rm el}^\sigma  
= \frac{1}{v_F} \int {\frac{d {\bf p}'}{(2\pi)^3}}
2\pi \; \delta(\xi'-\xi) |\langle {\bf p}' \sigma | {\cal T} | {\bf p}
\sigma\rangle|^2 \;,
\label{eq:sigma_el}
\ee
with $\xi$ the energy of the electron measured from the Fermi surface.
In contrast to $\sigma_{\rm el}^\sigma  $, the total
scattering cross section  also includes those scattering 
processes, where some excitations are left  behind, and thus the 
outgoing state is not a single particle state. The inelastic scattering 
cross section associated with these processes is thus clearly 
the {\em difference} of these two cross-sections:
\begin{equation}
\sigma_{\rm inel}^\sigma = \sigma_{\rm total}^\sigma - \sigma_{\rm
  el}^\sigma \; . 
\label{eq:difference}
\end{equation}
These processes are schematically shown in Fig.~\ref{fig:ineastic_elastic}

For quantum impurities in a free electron gas  it is more transparent to introduce
angular momentum channels, $L\equiv(l,m)$, and define the scattering states in 
terms of radially propagating states $|{\bf p},\sigma\rangle \to   ||{\bf
  p}|,L,\sigma \rangle$:
\be
|{\bf p},\sigma\rangle = \sqrt{\frac 2 \varrho }\sum_L \int d\Omega_{\hat p}
Y_L({\hat p})  ||{\bf
  p}|,L,\sigma \rangle\;,
\ee
with $Y_L({\hat p})$ the spherical functions, and $\varrho = 
\varrho(|{\bf  p}|)$ the density of states of the conduction electrons.
In this basis the on-shell $S$ and $T$-matrices become  matrices in the angular momentum 
quantum numbers that depend only on the energy $\omega\equiv\xi({\bf p},\sigma)$ of 
the incoming particle. The $S$-matrix can 
then
be expanded as 
\be
\langle \vp \sigma |{\cal S}|\vp'\sigma'\rangle  =  {\frac 2 \varrho }
\sum_{L,L'} Y^*_L({\hat \vp}) \; s_{L\sigma,L'\sigma'}(\omega)\; Y_{L'}({\hat \vp}')
\;, 
\ee
 and the on-shell $T$-matrix is given by a similar expression, the
 coefficients of the expansions being related by
\be
{s}_ {L,\sigma;\; L'\sigma'}(\omega) = \delta_{L,L'}\delta_{\sigma,\sigma'}  + i \; {t}_ {L,\sigma;\; L'\sigma'}(\omega)
\;. 
\ee
The eigenvalues ${\cal S}_\lambda$ of the matrix $s_ {L,\sigma;\; L'\sigma'}$  
must all be within the complex unit circle for any $\omega$, and they 
are directly related to the inelastic scattering cross section. 
To see this, let us consider the case of scattering only in the $s$-channel
($L=0$), and
assume spin conservation. In this case $s_ {L,\sigma;\; L'\sigma'}$ becomes a simple 
number, $s(\omega)=1+it(\omega)$, 
$$  
s(\omega) =
2\pi \varrho\; \langle \vp \sigma |{\cal S}|\vp'\sigma'\rangle = 
2\pi \varrho\; \;{\cal S}(\omega) 
$$
and the inelastic scattering cross section  can be expressed as 
\bea
\sigma_{\rm inel}(\omega)&=&\frac{\pi}{p_F^2}\;(1-|{s}(\omega)|^2)
\nonumber\\
&=&\frac{\pi}{p_F^2}\;(2\;{\rm Im}\;t(\omega)-|t(\omega)|^2)\;,
\label{eq:sigma_inel}
\eea
where we assumed  free electrons of dispersion $\xi={\bf p}^2/2m-\mu$
with  a Fermi energy $\mu$ and a corresponding Fermi momentum $p_F$.
Eq.~(\ref{eq:sigma_inel}) implies that 
 the scattering becomes {\em totally elastic} whenever ${s}(\omega)$ is on
the unit circle, and it is {\em maximally inelastic} if the corresponding 
single particle matrix element of the $S$-matrix vanishes.  
The former situation occurs at Fermi
liquid fixed points, while the latter case is realized, e.g., in 
the
two-channel Kondo model or the two-impurity Kondo model.
The total scattering cross section, on the other hand, is related 
to the real part of ${s}(\omega)$ as
\be
\sigma_{\rm tot}(\omega)= \frac{2\pi}{p_F^2}\;(1- {\rm Re}\{{s}(\omega)\})
=\frac{2\pi}{p_F^2}\;{\rm Im}\;t(\omega)\;.
\ee
It is easy to generalize this result  to the case of many
scattering channels, and one finds that inelastic scattering can take place 
only  if some of the eigenvalues 
of $s_ {L,\sigma;\; L'\sigma'}$ are not on the unit circle.~\cite{singular} 

To compute the off-diagonal matrix element 
$\langle {\bf p} \sigma | {\cal T} | {\bf p}' \sigma'\rangle\;,$ 
we first relate it 
to the conduction electrons' Green function through the so-called reduction  
formula detailed in Section~\ref{sec:reduction},~\cite{Itzykson80} 
\begin{eqnarray}
\label{eq:T-matrix}
&&\; \lefteqn{\langle  {\bf p},\sigma| {\cal T} | {\bf p}'  \sigma'\rangle =
  } \\
& & -  \; [G^0]^{-1}_{\pm{\bf p},\pm\sigma}(\xi)\; 
G_{\pm\vp \;\pm\sigma, \pm  \vp'\;\pm \sigma'} (\xi) \; 
[G^0]^{-1}_{\pm {\bf p}'\pm \sigma'}(\xi) 
\;.
\nonumber
\end{eqnarray}
Here the $\pm$ signs correspond to electron and hole states
with $\xi > 0$ and  $\xi < 0$ and of energy $E=|\xi|$, 
$G_0$ denotes the free electron Green's
function, and $G$ the full many-body time-ordered electron Green's function. 
Thus the positive frequency part of the Green's function describes the
scattering of electrons, while the negative frequency part that of holes. 
Strictly speaking, our derivation of this formula only holds for Fermi
liquids, i.e., for models where the ground state has no internal degeneracy
and can continuously be related to a non-interacting ground state. 
The consideration of internal ground state 
degeneracy needs some care, and the definition of inelastic scattering in that
case is not straightforward.~\cite{singular} However, the finite temperature 
results of Ref.~\onlinecite{Rosch} show that  Eqs.~(\ref{eq:sigma_tot}), 
(\ref{eq:sigma_el}), and (\ref{eq:difference}) together with 
(\ref{eq:T-matrix})
provide the physically meaningful definition even in this case at $T=0$.

According to Eq.~(\ref{eq:T-matrix}),  to compute the inelastic and elastic 
scattering cross-sections, we need to evaluate the self-energy of the conduction
electron's Green function. We do this by relating the self-energy to some local
correlation function, that we then compute 
either analytically within some approximation, or
numerically using the powerful
machinery of numerical renormalization group (NRG).~\cite{NRG_ref}
This step depends on the  specific impurity model at hand, and can be achieved
through equation of motion methods,~\cite{Costi} diagrammatically,~\cite{zar_inel}
or through a straightforward path integral treatment, as we do it in 
Section~\ref{appen:tmatrixkondo}. The recently-formulated scattering
Bethe ansatz approach can possibly provide a way to avoid this
numerical computation, and determine the full energy-dependence of the  
$S$-matrix analytically~\cite{mehta}.

\begin{figure}
\includegraphics[width=0.9\columnwidth,clip]{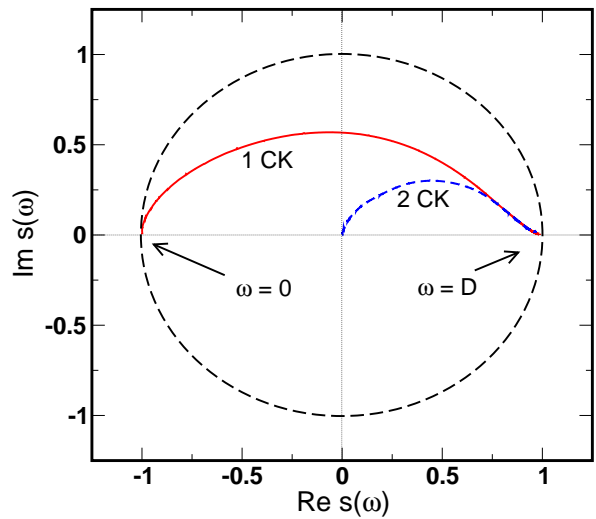}
\caption{\label{fig:Sflow}
Renormalization group flow of the eigenvalues of the single 
particle $S$-matrix for the single- and two-channel Kondo
models. 
For the single-channel Kondo model the scattering becomes 
elastic ($|S|=1$)  both in the limit of very large
and very small energies,  $\omega\to \infty$, and  $\omega\to 0$, respectively. 
For the 2-channel Kondo model ${s}(\omega \to 0) =0$, implying that half
of the scattering remains inelastic even for $\omega\to 0$.
}
\end{figure}

Let us close this Introduction by  illustrating  the power of this 
formalism through the simple examples of the single- and two-channel 
Kondo models, defined through the Hamiltonian:
\bea
H&=&
\sum_{\alpha=1}^f \sum_{{\bf p}, \sigma}  
\xi_{\bf p}\; a^{\dagger}_{{\bf p} \sigma,\alpha}a_{{\bf p} \sigma,\alpha}
\nonumber \\
&+& \frac J2  \vec{S} 
\sum_{\alpha=1}^f   
\sum_{\genfrac{}{}{0pt}{}{{\bf p},{\bf p}'}{\sigma\sigma'}}
a^{\dagger}_{{\bf p}\sigma,\alpha} {\vec{\sigma}}_{\sigma\sigma'} a_{{\bf p}'\sigma',\alpha} 
\;.
\label{Eq:Kondo}
\eea
Here  $a^\dagger_{{\bf p}\sigma\alpha}$ creates a conduction electron with
momentum  $\bf p$, spin $\sigma$ in channel $\alpha$, $S=1/2$ is the impurity
spin, and $f=1$ and $f=2$ for the single- and two-channel
Kondo models, respectively.

In both models the $T$-matrix can be related to
the Green's function of the so-called composite Fermion operator, 
$F_{\sigma\alpha} \equiv \sum_{\sigma', {\bf p}} 
{\vec S} \cdot {\vec \sigma}_{\sigma\sigma'} a_{{\bf p}
  \sigma'\;\alpha}$,~\cite{Costi}
which can then be computed using NRG.~\cite{NRG_ref} 
The evolution of the eigenvalue of the numerically obtained $S$-matrix 
is shown in Fig.~\ref{fig:Sflow}.  
In both cases, ${s}(\omega) = {s}(\omega/T_K)$ is a universal function 
that depends only on the ratio $\omega/T_K$, with $T_K$ the so-called Kondo
temperature,
$T_K\approx E_F\; e^{-1/\varrho J}$, with $E_F$ the Fermi energy
and $\varrho$ the density of states at the Fermi energy for one spin 
direction.~\cite{Hewson,Cox_Zawa}

For the single-channel Kondo model the scattering becomes 
elastic both in the limit of very large and 
very small energies,  $\omega\gg T_K$, and  $\omega\ll T_K$, respectively, 
where the eigenvalues lie on the unit circle. 
The reasons are different: At large energies conduction electrons 
do not interact with the impurity spin efficiently. 
At very small energies, on the other hand, the impurity's spin is screened and
disappears from the problem apart from a residual phase shift of 
$\pi/2$ and an irrelevant local electron-electron interaction.~\cite{Nozieres}
The maximum inelastic scattering
is reached when the eigenvalue ${s}(\omega)$ is  closest to the
origin, i.e., at energies in the range of the Kondo temperature,
$\omega\approx T_K$.

For the 2-channel Kondo model, on the other hand,  ${s}(\omega \to 0) =0$, 
implying that the scattering is maximally inelastic even at the Fermi energy, $\omega\to 0$.
This property of the $S$-matrix has been 
first noticed by  Maldacena and Ludwig\cite{Maldacena}, 
and is characteristic of a non-Fermi liquid, where incoming 
electrons do not scatter into an outgoing single electron state, even at the
Fermi energy. Note, however, that the vanishing of the $S$-matrix does not
imply that all the scattering is fully inelastic. In fact, from 
Eqs.~(\ref{eq:sigma_tot}) and (\ref{eq:sigma_el}) it follows that 
$$
\sigma_{\rm inel}^{\rm 2CK}=\sigma_{\rm el}^{\rm 2CK}=
\frac {\sigma_{\rm tot}^{\rm 2CK}}2\;,
$$
i.e., half of the scattering remains still elastic. In other words, in the
elastic channel, the
unscattered and scattered $s$-electron wave functions are completely out of
phase, and therefore there is no net outgoing particle 
in the $s$-channel.
\begin{figure}
\includegraphics[width=7cm,clip]{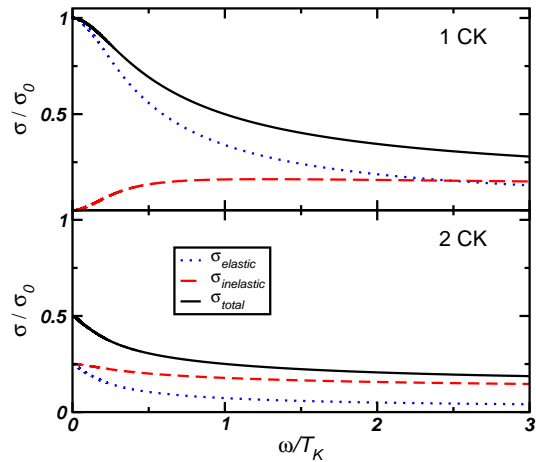}
\caption{
\label{fig:sigma_summary}
Top: Inelastic, elastic, and
  total scattering 
  rates
  for the single-channel Kondo model
  in units of $\sigma_0 = 4\pi/p_F^2$ for a spin up electron.
  $\sigma_{\rm inel}$
  scales approximately linearly with $\omega$ for $0.05\;T_K <
  \omega < 0.5\;T_K$, behaves as $\sim \omega^2$ for $\omega < 0.05\;T_K$, 
  and remarkably, shows a plateau above $T_K$.
Bottom: Same for the two-channel Kondo problem for a spin up electron in
channel 1. Note that the inelastic scattering cross section remains finite 
at $\omega=0$, and is precisely half
of the elastic scattering cross section.} 
\end{figure}

The structure of the flow of ${s}(\omega)$ appears more directly in the 
in the energy dependence of the various scattering cross-sections shown in 
Fig.~\ref{fig:sigma_summary} for these two cases. In the single-channel case, it
is quite remarkable that the 
low-energy $\sim \omega^2 $ inelastic scattering cross section 
expected from Fermi liquid considerations\cite{Nozieres} is limited
 only to the regime $\omega < 0.05\;T_K$, and for $0.05\;T_K <
\omega < 0.5\;T_K$ the inelastic scattering cross section is 
quasi-linear. Furthermore, 
above $T_K$ a wide plateau appears (rather then a peak), where   
 $\sigma(\omega)$  is large and almost independent of the energy $\omega$ of
 the incoming particle. Both features appear also in a finite temperature 
calculation,~\cite{Rosch}  are in quantitative agreement with the experimental 
results of Refs.~\onlinecite{Saminadayar2006,mohanty2003} 
on magnetically doped 
wires, and provide a possible explanation of the observed
saturation of the dephasing time in some experiments on
dephasing~\cite{Mohanty-Webb}. 
As we show in Section~\ref{sec:inelAnderson}, these universal features
are robust and present in the Anderson model too.

It is important to emphasize here that in the present paper
we computed the inelastic scattering rate of {\em electrons}, rather
than that of {\em quasiparticles}. This is motivated by the trivial observation that 
in a real experiment, the external electromagnetic field couples with a
minimal coupling to the bare conduction electrons. 
Precisely for this reason, the Kubo formula contains the  conduction electron 
current operators, and also, the relevant quantity to
determine dephasing  is thus  the inelastic scattering rate 
 of  electrons. This is what we have computed here and that has been computed 
 in Refs.~\cite{zar_inel} and \cite{Rosch}.  

The definition of {\em quasiparticles} depends on the context in which they emerge.
If one defines them  as {\em stable} elementary excitations of the vacuum, as Nozi\`eres 
did \cite{Nozieres}, or as they appear in Bethe ansatz~\cite{BA} , then, by definition, 
these quasiparticles do not decay at all at $T=0$ and scatter only 
elastically \cite{Nozieres}.

Such quasiparticles are, however, usually complicated  objects in terms of conduction
electrons. For this reason they are  typically not minimally 
coupled to the gauge field, and therefore, the current operator in the 
Kubo formula is a very complicated many-body vertex in the language of 
quasiparticles. 
Excepting for $\omega=0$, a real conduction electron is composed of 
 many such  stable quasiparticles, and it  decays inelastically even at
 $T=0$  temperature, even if quasiparticles do not. In the Kondo model, at the Fermi
energy quasiparticle states are simple
phase shifted conduction electron states. However, the connection between 
quasiparticles and conduction electrons is not trivial for any finite energy.
Therefore, if one considers inelastic scattering at a finite energy, one must
{\em precisely} specify  
how finite energy quasiparticle states are defined, how they couple to a gauge
field, and how a finite energy electronic state is decomposed in terms of
these quasiparticles. Unfortunately, except for the Bethe ansatz, we are not aware of any work 
which would provide this necessary connection in sufficient detail,
and would go beyond a simple heuristic treatment (which might still give
the correct result). In the present framework, we avoided  this difficulty by
formulating the problem in terms of electrons rather than quasiparticles.

The paper is organized as follows:
In Section \ref{sec:reduction} we present the
derivation of
the reduction formulas.
In Section~\ref{appen:tmatrixkondo}
we determine the T-matrix for the Kondo model.
In Sections~\ref{sec:inel1CK}, \ref{sec:inel2CK}, and
\ref{sec:inelAnderson} we present results on the inelastic scattering
rate for the single- and two-channel Kondo model and for the Anderson model,
respectively. In Section~\ref{sec:conclusion} the results are summarized.
In the Appendix some details of the
derivation of the T-matrix 
for the Anderson model is discussed.

\section{Reduction Formulas}
\label{sec:reduction}

\subsection{Definition of scattering states in the Heisenberg picture}

Although reduction formulas are often used in the literature in a heuristic
way, apart from the derivation of Langreth 
for the Anderson model,~\cite{langreth}  
we do not know of any work that would establish a rigorous connection 
between the single-particle matrix elements of the $T$-matrix 
and the conduction electron's Green's function for a general quantum 
impurity problem. Here we therefore present a short derivation of the
reduction formulas by generalizing the  procedure used in the domain of 
particle physics.~\cite{Itzykson80,LeBellac}

In this section, following the field theoretical language, we 
shall use the Heisenberg picture, and describe scattering in terms of 
the field operators,~\cite{Itzykson80} $\psi_\sigma(x)$, where 
we introduced the four-vector notation, $x \equiv (t, \vx)$.
The evolution of this field operator is described by the time-dependent 
Hamiltonian, with the interactions switched on and off adiabatically with a rate 
$\delta \to 0$,
\be 
H \equiv H(t) = H_0 + e^{-\delta |t|} H_{\rm int}\,.
\ee
Here $H_0$ denotes the non-interacting Hamiltonian, 
\be 
H_0 = \int d^3\vx \; \psi^\dagger_\sigma(x)[ -\frac{\Delta}{2m}-\mu]\psi_\sigma(x), 
\ee
with $\Delta$ the Laplace operator, and $\mu$ the chemical potential of the
electrons. 
The interaction part $H_{\rm int}$ does not need to be specified 
at this point, and depends on the particular model considered. 
For the sake of simplicity we assume that the quantum impurity interacts
with free electrons, but the procedure described can be generalized 
for electrons with more complicated dispersions, too. 

Within the Heisenberg picture, states are independent of time, and all
non-trivial scattering is incorporated in the time evolution of the 
fields. Scattering states can be defined through the asymptotic 
form of the field operators. Incoming and outgoing scattering states 
can be defined based on the simple observation that for times $t\to \pm \infty$
the equation of motion of $\psi_\sigma(x)$ is generated by $H_0$, and
therefore $\psi_\sigma(x)$ behaves asymptotically as a free field:
\be
\psi_\sigma(x,t\to -\infty) \to  \int \frac{d^3 \bf p}{(2\pi)^3}\,e^{-i \,p\cdot x}
a_{p \,\sigma,\textrm{in}}\,,  
\ee
where $a_{p \,\sigma,\textrm{in}} \equiv a_{\vp \,\sigma,\textrm{in}}  $ are 
just the annihilation operators of incoming (one particle) scattering states. 
Here for the sake of compactness, we introduced the four-momentum,  $p \equiv (\xi, \vp)$, 
with $\xi = \xi({\bf p}) = {\bf p}^2/2m - \mu$ the energy of the conduction
electrons, measured from the Fermi energy, and 
$p\cdot x =  \xi \; t - \vp \; \vx$.
The operators 
$a_{p \,\sigma,\textrm{in}} \equiv a_{\vp \,\sigma,\textrm{in}}  $ satisfy
standard anti-commutation relations:
\be
\{a_{p \,\sigma,\textrm{in}}, a^\dagger_{p' \,\sigma',\textrm{in}}\}
= (2\pi)^3 \; \delta_{\sigma,\sigma'}\; \delta(\vp - \vp')\;.
\ee
Note that the operators $a^\dagger_{p\,\sigma,\textrm{in}}$
do not create free electrons, rather, they are creation operators of incoming 
electrons in {\em scattering states}, which are {\em asymptotically free}. 
\footnote{Here we make use of the fact that the $Z$-factor for a quantum
impurity problem in the infinite volume limit is $Z=1$.}

The operators $a^\dagger_{p\,\sigma,\textrm{in}}$ can be used to construct  
incoming single particle scattering states, 
$|\vp,\sigma \rangle_{\textrm{in}}$.
For electrons,  i.e., excitations of 
momenta larger than the Fermi momentum, $|\vp|> p_F$, these scattering states 
can be simply defined as\footnote{Here we used the fact the incoming and
outgoing vacuum states are isomorphic, and denoted both of them by $|0\rangle$. }
\be
|\vp,\sigma \rangle_{\textrm{in}} \equiv
a^\dagger_{p,\sigma,\textrm{in}}|0\rangle 
=
\lim_{t \to -\infty} \int \textrm{d}^3{\vx} \; e^{-ip\cdot x}\; \psi^\dagger_{\sigma}(x)
|0\rangle \;.
\label{eq:instate}
\ee
Outgoing single electron scattering states, 
$|p,\sigma \rangle_{\textrm{out}}\equiv |\vp,\sigma \rangle_{\textrm{out}}$,
can be defined in a similar way, by  expanding
the field $\psi_\sigma(x,t\to +\infty)$,  
\begin{equation}
|\vp,\sigma \rangle_{\textrm{out}} \equiv
a^\dagger_{p,\sigma,\textrm{out}}|0\rangle 
=
\lim_{t \to +\infty} \int \textrm{d}^3{\vx} \; e^{-ip\cdot x}\; \psi^\dagger_{\sigma}(x)
|0\rangle \;.
\label{eq:outstate}
\end{equation}

Incoming and outgoing hole states must be defined slightly differently, 
because an incoming hole of energy $E>0$, momentum 
$\vp$, and  spin $\sigma$ is created by {\em removing} an electron of energy 
$\xi=-E$, momentum $-\vp$, and spin $-\sigma$ from the Fermi surface.
In other words, incoming hole scattering states are defined for $|\vp| < p_F$ as 
\bea
&&|\vp,\sigma \rangle_{\textrm{in/out}}\equiv
a_{-p,-\sigma,\textrm{in/out}}|0\rangle 
\label{eq:hinstate}
\\
&&\phantom{aa}=\lim_{t \to \mp \infty} \int \textrm{d}^3{\vx} \; e^{-ip\cdot x}\; \psi_{-\sigma}(x)|0\rangle \;
\phantom{aaaa}(|\vp| < p_F)\;.
\nonumber
\eea

\subsection{Reduction formulas and Green's functions}
\label{appen:reductionformulasforelectrons}

We proceed to derive a general relation involving Green's functions to express
the off-diagonal matrix elements  
\begin{eqnarray}
i \; \langle \vp\;\sigma|\hat{T}|\vp'\sigma' \rangle = \me \quad \quad (\vp\neq\vp')  
\label{equ:delta}
\end{eqnarray}
for electronic excitations with $|\vp| > p_F$  first.
Using the asymptotic expression Eq.~(\ref{eq:instate}) this matrix element 
can be expressed as
\begin{eqnarray}
\me &=& \nonumber \\ &&\hspace*{-3cm} \lim_{y_0 \to
  -\infty}\phantom{a}_{\textrm{out}} \langle \vp,\sigma | \int_{y_0}
\textrm{d}^3\vy \; e^{-ip'y} \psi^\dagger_{\sigma'}(y) |0\rangle \; .
\end{eqnarray}
Integrating by part we obtain 
\begin{eqnarray}
&&\lim_{y_0 \to -\infty} \int_{y_0} \textrm{d}^3\vy \; e^{-ip'y}
\psi^{\dagger}_{\sigma'}(y)=\nonumber 
\\ &=&-\int \textrm{d}^4y \frac{\partial}{\partial y_0}\left[e^{-ip'y}
  \psi^\dagger_{\sigma'}(y)\right]
+a^\dagger_{p',\sigma',\textrm{out}}.
\nonumber  
\end{eqnarray}
The last term does not give a contribution to the matrix element 
for $p\ne p'$, therefore we drop it. The rest can be expressed 
as
\begin{eqnarray}
&&
\int_{-\infty}^{\infty} \textrm{d}^4y 
\frac{\partial}{\partial y_0} \left[e^{-ip'y} \psi^\dagger_{\sigma'} (y)
\right]
=\nonumber 
\\
&&- i\int \textrm{d}^4y e^{-ip'y} \left[i\frac{\partial}{\partial y_0}+H_0(\vy )
\right] \psi^\dagger_{\sigma'}(y), 
\nonumber  
\end{eqnarray}
where we obtained the r.h.s. of this equation by using the fact that 
$p'$ is  on the energy shell, and therefore 
$p_0' e^{-ip'y}= (-\frac 1 {2m} \Delta_{\bf y} - \mu)e^{-ip'y}$, and then
by integrating by part with respect to ${\bf y}$.
Thus the off-diagonal matrix elements of the 
$S$-matrix simplify to 
\begin{eqnarray}
&&
\me
=
\\
&&\phantom{aaaa}i\int \textrm{d}^4 y e^{-ip'y} \left[i\frac{\partial}{\partial
    y_0} + H_0(\vy) \right]
\phantom{a}_{\textrm{out}}\langle \vp
\sigma|\psi^{\dagger}_{\sigma'}(y)|0\rangle \nonumber 
\label{equ:equ1}
\end{eqnarray}

Using the  asymptotic relation of the outgoing states, \eqref{eq:outstate},
 we can now write the  full matrix element as
\begin{eqnarray}
&&\me =i \int \textrm{d}^4y e^{-ip'y} \left[i \frac{\partial}{\partial
    y_0}+H_0(\vy)\right] \times \nonumber \\ &\times&\lim_{x_0\to \infty} \int
\textrm{d}^3\vx \; e^{ipx} \langle0|\psi^{\phantom{\dagger}}_{\sigma}(x)\psi^{\dagger}_{\sigma'}(y)|0\rangle.
\end{eqnarray}
Once again, we convert the last integral into an integral over the 
whole space-time, which yields
\begin{eqnarray}
&& \lim_{x_0\to \infty} \int \textrm{d}^3\vx\; 
e^{ipx}
\langle0|\psi^{\phantom{\dagger}}_{\sigma}(x)\psi^{\dagger}_{\sigma'}(y)|0\rangle
\nonumber \\  &=& \int^{\infty}_{-\infty} \textrm{d}^4{x}
\frac{\partial}{\partial x_0} e^{ipx} \langle0|T
\psi^{\phantom{\dagger}}_{\sigma}(x)\psi^{\dagger}_{\sigma'}(y)|0\rangle
\end{eqnarray} 
where the time ordering operator $T$ has been inserted 
to assure that the $x_0 \to -\infty$ contribution vanishes
by Eq.~(\ref{eq:instate}): 
$\lim_{x_0 \to -\infty} \int \textrm{d}^3\vx \; e^{ipx}
\langle0|\psi^{\dagger}_{\sigma'}(y)\psi^{\phantom{\dagger}}_{\sigma}(x)|0\rangle
= 0$.

We can manipulate the remaining expression in the same way as before
to finally obtain 
\begin{widetext}
\begin{eqnarray}
&& \me=-\int \textrm{d}^4x \;\textrm{d}^4y\; 
e^{-ip'y}e^{ipx}
\left[
\left(-i\;   \overrightarrow{\partial}_{x_0}-\vec H_0(\vx)\right) 
\langle \textrm{T}\psi^{\phantom{\dagger}}_{\sigma}
(x)\psi^{\dagger}_{\sigma'}(y)\rangle
\left(i\;  \overleftarrow{\partial}_{y_0}-H_0(\vy)\right) 
\right]\;,  
\label{final}
\end{eqnarray}
\end{widetext}
where the arrows indicate forward and backward differentiation, respectively. 
Observing that the operator  $(-i\; { \partial}_{x_0}- H_0(\vx)) \;\delta(x-x')$ 
is simply the matrix element of the inverse  of the non-interacting 
Green's function, 
\bea
\langle x'| \hat G^{-1}_0 |  x\rangle &=& \delta(x'-x)\;
\left(-i\; \overrightarrow{ \partial}_{x_0}- H_0(\vx)\right) \;,
\nonumber
\\
&=& 
\left(i\; \overleftarrow{ \partial}_{{x}_0'}- H_0(\vx')\right)\;\delta(x'-x)\;,
\nonumber
\eea 
Eq.~(\ref{final}) can be simply expressed as
\begin{eqnarray}
&&\me =
\label{eq:conv}
\\
&&\phantom{aaa}-i \int \textrm{d}^4x \;\textrm{d}^4y\; 
e^{ipx -ip'y}
\left[ \hat G^{-1}_0 \ast  \hat G \ast \hat G^{-1}_0\right]_{x,y}\;, 
 \nonumber 
\end{eqnarray}
with '$\ast$' the four-dimensional convolution operator,  and $\hat G$ the
usual interacting Green's function, 
\be 
 G_{\sigma \sigma'}(x,y) = -i\;\langle T\psi^{\phantom{\dagger}}_{\sigma}
(x)\psi^{\dagger}_{\sigma'}(y)\rangle\;.
\ee
The Fourier transformation of \eqref{eq:conv} then yields
\begin{eqnarray}
&& \me= 
\label{me:electron}
\\
&& \phantom{a}-i \;
G^{-1}_{0,\sigma}(p)G^{\phantom{1}}_{\sigma,\sigma'}(p,p')G^{-1}_{0,\sigma}(p').
\phantom{aaaaaa} (|\vp|>p_F)
\nonumber  
\end{eqnarray}
Translational invariance in time further implies 
\begin{eqnarray}
G^{\phantom{1}}_{\sigma,\sigma'}(p,p')
&=& \phantom{a} \;2 \pi \;\delta (\xi (\vp')-\xi(\vp)) \;
G^{\phantom{1}}_{ \vp\sigma, \vp'\sigma'}(\xi(\vp)).
\nonumber
\end{eqnarray}
Inserting this into \eqref{me:electron} and
 comparing it with  Eq.~(\ref{eq:on_shellT}) yields 
Eq.~(\ref{eq:T-matrix}) for $\xi>0$. 

The derivation for holes follows exactly the same lines 
excepting that the matrix element to be computed is now 
\begin{eqnarray}
\me
=\langle 0|a^{\dagger}_{-p,-\sigma,\textrm{out}} a^{\phantom{\dagger}}_{-p',-\sigma',\textrm{in}}|0\rangle,
\label{equ:maelement2}
\end{eqnarray}
and correspondingly, the final expression of the $S$-matrix element now reads
for $\vp\ne \vp'$
\begin{eqnarray}
&& \me= \phantom{aaaaaaaaaaaaaaaaaa} (|\vp|<p_F)
\nonumber  
\\
&& \phantom{aaaa}i \;
G^{-1}_{0,-\sigma'}(-p')G^{\phantom{1}}_{-\sigma',-\sigma}(-p',-p)G^{-1}_{0,-\sigma}(-p)\;.\nonumber \\
\nonumber  
\end{eqnarray}

%


\section{T-matrix of  the Kondo model}

\label{appen:tmatrixkondo}

For practical calculations, one needs to determine the $T$-matrix of
Eq.~(\ref{eq:T-matrix}) somehow. For most really interesting cases this can be
done analytically only approximately and in a limited energy range, and
numerical methods must be used.  The most adequate way to perform the
calculation is to  first relate the  $T$-matrix to some local correlation
function that can then be computed using Wilson's numerical renormalization
group method.~\cite{NRG_ref} To establish the desired relation, one can use 
equation of motion methods,~\cite{Costi} or do diagrammatic perturbation
theory and sum up the diagrams up to infinite order,~\cite{zar_inel} 
but here we show yet another rather elegant way, in terms of path integrals. 

\begin{figure}
\begin{center}
\includegraphics[width=8cm,clip]{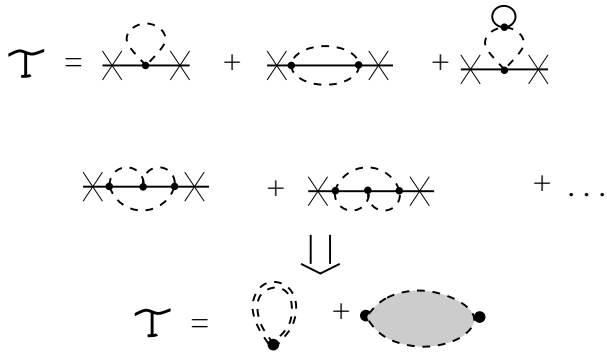}
\end{center}
\caption{
Diagrammatic
  representation of the $T$-matrix in the Kondo problem. Dashed lines
  denote pseudo-fermion propagators and describe the evolution of the
  impurity spin, while continuous lines denote free conduction
  electron propagators.  Filled circles stand for the exchange
  interaction $J$.  The first term of the $T$-matrix is simply
  proportional to the expectation value of the impurity spin, it is
  frequency independent, and vanishes in the absence of magnetic
  field.  The second term can be identified as the composite fermions
  correlation function.
}
\label{tmatrixdiagrams}
\vspace{0.5cm}
\end{figure}

Although this method works for essentially any quantum impurity problem, here
we show how it works for the Kondo model, already  defined in the introduction (see
Eq.~(\ref{Eq:Kondo})). 
The application of this method to the Anderson model is discussed in the
Appendix.
To use a field theoretical formalism, following
Abrikosov, we represent the impurity spin $\vec S$ by fermionic operators, $f_{\sigma}$,
\begin{eqnarray}
\vec{S}=\frac 12
\sum_{\sigma,\sigma'}f^{\dagger}_{\sigma}\vec{\sigma}_{\sigma\sigma'}f^{\phantom{\dagger}}_{\sigma'}\;,
\phantom{nnn} 
\sum_{\sigma}f^{\dagger}_{\sigma}f^{\phantom{\dagger}}_{\sigma} \equiv 1\;.
\end{eqnarray}
Next, we define the generating functional for the conduction electron Green's
functions as follows: 
\be
Z[\eta_\sigma, \overline {\eta}_\sigma] = \int D \left[a_{\sigma} 
\overline{a}_{ \sigma} \right]
\tilde D \left[ 
f_{\sigma} \overline{f}_{\sigma}  \right]e^{-i\;S}e^{i 
\overline
{\eta}\cdot a + i \overline a
\cdot  {\eta}},
\label{pathint}
\ee
where the tilde on the second integration measure indicates that 
one must impose the constraint  $\sum_{\sigma}f^{\dagger}_{\sigma}f^{\phantom{\dagger}}_{\sigma} \equiv 1$
when performing the path integral, and
we introduced the shorthand notation:
$ 
\overline {\eta}\cdot a  \equiv
\sum_{{\bf   p},\sigma} \int dt\;  \overline {\eta}_{{\bf p},\sigma}(t) a_{{\bf p},\sigma}(t)\;.
$
The action $S$ in Eq.~(\ref{pathint}) consists of three terms,  
$S = S_e + S_f + S_{\rm int}$: The first term, $S_e$ describes the conduction
electrons, 
\be 
S_e = - \sum_{\vp,\sigma }\int dt\;dt'\; \overline a_{\vp \sigma}(t) [G^0]^{-1}_{\vp}(t-t')\; a_{\vp \sigma}(t') 
\label{eq:Se}
\ee
with $G^0$ the time-ordered free electron Green's function. The term 
$
S_f \equiv - i \; \sum_{\sigma }\int dt\; \overline f_{\sigma}(t)\; \frac{d}{dt}  f_{ \sigma}(t)
$
generates the spin dynamics, while the last term simply describes the
interaction:
\be
S_{\rm int} \equiv \frac J 2 \; \sum_{\genfrac{}{}{0pt}{}{\vp,\vp'}{\sigma,\sigma'}}
\int dt\; \vec S (t)\; 
\overline a_{\vp \sigma}(t) \vec \sigma_{\sigma\sigma'} a_{\vp' \sigma'}(t) \;.
\ee
The full time-ordered Green's function is related to $Z$ by
\begin{eqnarray}
G_{\vp \sigma,\vp'\sigma'}(t-t')
=\left.
\frac 1 i {\frac{\delta^2 \textrm{ln}Z[\overline{\eta},\eta]}
{\delta \overline{\eta}_{\vp \sigma}(t) \delta
\eta_{\vp'\sigma'}(t')}}\right|_{\eta =\overline{\eta}=0}\;.
\label{Eq.:greens}
\end{eqnarray}
We can  derive the required identity by simply shifting the integration variable in 
Eq.~(\ref{pathint}), 
\be 
a_{\vp \sigma}(t) \to a_{\vp \sigma}(t) - \int dt'\; G^0_{\vp}(t-t')\; \eta_{\vp \sigma}(t')\;. 
\ee
As a result, the exponent in Eq.~(\ref{pathint}) transforms to:
\begin{widetext}
\bea
S - \overline {\eta}\cdot a - \overline a \cdot  {\eta}  \phantom{a}
\Rightarrow \phantom{a}
 S &+& 
 \sum_{\vp, \sigma}
\int dt\;dt'\;  \overline \eta_{\vp \sigma}(t) 
G^0_{\vp\sigma }(t-t') \eta_{\vp \sigma}(t') \;
- \frac J 2 \sum_{\vp, \sigma}
\int dt\;dt'\; \left( \overline F_{ \sigma}(t) 
G^0_{\vp\sigma }(t-t') \eta_{\vp \sigma}(t') \; + {\rm h.c.}\right)
\nonumber
\\
&&
+
\frac J 2 \;
 \sum_{\genfrac{}{}{0pt}{}{\vp,\vp'}{\sigma,\sigma'}}
\int dt\;dt'\; dt''\; 
\overline \eta_{\vp \sigma}(t)
G^0_{\vp\sigma }(t-t') 
\vec S (t') \vec \sigma_{\sigma\sigma'}
G^0_{\vp\sigma }(t'-t'') \eta_{\vp' \sigma}(t'')
\nonumber
\;, 
\eea
where we introduced the composite fermion field, 
$F_\sigma(t)\equiv \sum_{\vp,\sigma'} \vec S(t) \vec \sigma_{\sigma\sigma'}
a_{\vp,\sigma'}(t)$. 
Carrying now out the functional derivation 
of \eqref{Eq.:greens}
we obtain the following simple relation
\be
G_{\vp \sigma,\vp'\sigma'}(t-t') = 
\delta_{\vp,\vp'}\;G^0_{\vp \sigma}(t-t')   
 + \int d\tilde t\;d\tilde t'\;
 G^0_{\vp \sigma}(t-\tilde t)  
\left(
\delta(\tilde t-\tilde t')\; 
\frac J 2 \langle \vec S\rangle \vec \sigma_{\sigma\sigma'}
 -i\; \frac{J^2}4 \langle 
F_\sigma(\tilde t) \overline F_{\sigma'}(\tilde t')\rangle 
\right)
 G^0_{\vp' \sigma'}(\tilde t'- t').
\label{Ward}
\ee
\end{widetext}
The average in this expression must be carried out by computing 
the appropriate  path integral, and results in the corresponding
time-ordered Green's function. 
Comparing Eqs.~(\ref{Ward}) and (\ref{eq:T-matrix}), and using the analytical 
properties of the time-ordered and retarded Green's functions at $T=0$
temperature, we finally obtain the relations:
\begin{widetext}
\bea
{\rm Im} \{ {\cal T}_{\tau \sigma,\tau\sigma'}(\omega)   \} &=&  \pi \frac {J^2}4
\varrho_{F,\sigma\sigma'}(\omega)\;, 
\label{spectral}
\\
-\tau \ {\rm Re}\{ {\cal T}_{\tau \sigma,\tau \sigma'}(\omega)\} &=&  
 \frac {J}2\langle \vec S\rangle \vec \sigma_{\sigma\sigma'}
+  \frac {J^2}4 \; {\cal P}\hspace{-.35cm}\int d\omega'\; \frac{\varrho_{F,\sigma\sigma'}(\omega)}{\omega-\omega'}
\;, 
\nonumber
\eea
\end{widetext}
with $\varrho_{F,\sigma\sigma'}(\omega)$ the spectral function of the composite
Fermion's Green function, and $\tau={\rm sgn}\; \xi$. 
Note that scattering takes place only in the $s$-channel, and therefore these
matrix elements do not depend  on the incoming and outgoing momenta of the excitations.
Eqs.~(\ref{Ward}) and (\ref{spectral}) can be easily
visualized in terms of diagrammatic perturbation theory, as shown in  
Fig.~\ref{tmatrixdiagrams}. The spectral function appearing in Eq.~(\ref{spectral})
is just a local correlation function that can be easily obtained through the
NRG method.~\cite{NRG_ref}    In the following  sections we shall 
primarily use this method 
to compute the single particle $T$-matrix and the inelastic scattering rates 
of the basic quantum impurity models, the single-channel Kondo model, the
two-channel Kondo model, and the Anderson model. Calculations for the spin
$S=1$ Anderson model have been performed in Ref.~\onlinecite{Hewson2}

\section{Inelastic scattering in the Kondo model}
\label{sec:inel1CK}
In the previous sections we have related the single particle $T$-matrix
and therefore the elastic and inelastic scattering amplitude of
electrons with {\em local} correlation functions through the reduction
formulas. 
In this section we shall use these results to analyze the $T=0$ temperature 
scattering properties of the Kondo model
using Wilson's NRG\cite{NRG_ref}. 
However, before presenting detailed numerical results, let us shortly discuss 
what one can learn from simple perturbation theory. 

Let us discuss the high-energy scattering of conduction electrons in the
absence of external magnetic field. In this 
limit one can attempt to do perturbation theory, and in first non-vanishing 
order one obtains
\be
t(\omega) = i \;2\pi \varrho \; {\cal T}(\omega) = -i
\frac{\pi^2}2\; \;S(S+1)\;  j ^2+\dots\;, 
\ee
where the dimensionless coupling $j=\varrho J$ has been introduced.
Summing up the leading logarithmic diagrams
amounts in
replacing $J$ by the renormalized coupling,  and gives
$$
t(\omega\gg T_K) \approx   i \frac{\pi^2}2
S(S+1) \frac 1 {\ln^2(\omega/T_K)}\;,
$$
with $T_K \sim E_F e^{-1/J\varrho }$ the Kondo temperature.
Thus, in leading logarithmic order, the total scattering cross section 
is given by:
$$
\sigma_{\rm tot}(\omega\gg T_K) \approx   \frac{\pi^3}{p_F^2} \;
S(S+1)  \frac 1 {\ln^2(\omega/T_K)}\;.
$$
The first non-vanishing contribution to the elastic scattering cross section, 
on the other hand, scales as $\sigma_{\rm el}\sim |t|^2\sim j^4$, and 
therefore $\sigma_{\rm el}(\omega)$ asymptotically behaves as
$$
\sigma_{\rm el}(\omega\gg T_K) \approx \frac{\pi^5}{4\;p_F^2} \;
S^2(S+1)^2  \frac 1 {\ln^4(\omega/T_K)}\;. 
$$ 
This implies that asymptotically, all the scattering is {\em inelastic}
\be
\sigma_{\rm inel}(\omega\gg T_K) \approx \sigma_{\rm tot}(\omega)
\approx \frac{\pi^3}{p_F^2} \;
S(S+1)  \frac 1 {\ln^2(\omega/T_K)}\;.
\label{sigma_inel_asymp}
\ee

This is a very surprising result, and contradicts to the conventional wisdom,
which tries to associate inelastic scattering with spin-flip scattering
from a free spin. In fact, this rather non-trivial result has been 
explained in Ref.~\onlinecite{Garst} in the following way: At high energies, 
incoming electrons are scattered 
by  the impurity spin fluctuations. These fluctuations can absorb an energy of 
the order of $\sim   T_K$, and therefore the energy of the incoming electron
is not conserved in leading order, but it typically changes by a tiny 
amount, $ \delta\omega \sim T_K$. In 
the most pedestrian perturbative approach
this tiny energy 
transfer 
is
neglected
and 
therefore one concludes incorrectly that the energy is conserved 
in leading order.

We can also
relate the cross sections above to scattering rates. 
Assuming a finite concentration $n_{\rm imp}$ of magnetic impurities, we can
compute the impurity averaged conduction electron Green's functions
and from that the conduction electron lifetime:
\be 
\frac {1}{\tau} 
= n_{\rm imp}\; v_F \; \sigma_{\rm tot}(\omega)
\approx n_{\rm imp} 
\frac {  \pi  S(S+1) } {2\; \varrho \;\ln^2(\omega/T_K)}\;.
\ee
In fact, the first part of this equation gives a general 
rule to connect various cross sections to the corresponding scattering 
times, and for the inelastic scattering rate, e.g., we have 
\be 
\frac 1 {\tau_{\rm inel}} 
= n_{\rm imp}\; v_F \; \sigma_{\rm inel}(\omega)\;.
\ee
For very large frequencies, again, the inelastic scattering 
rate is approximately equal to the elastic scattering rate:
\be 
\frac 1 {\tau_{\rm inel}} \approx n_{\rm imp}  
\frac {  \pi  S(S+1) } {2\;\varrho \;\ln^2(\omega/T_K)}\;.
\ee  
Note that this rate is a factor of 3/2 larger than the Nagaoka-Suhl
expression, which only takes into account spin flip processes.~\cite{Nagaoka}

For energies $|\omega|\ll T_K$ perturbation theory breaks down, and it is more
appropriate to use Nozieres' Fermi liquid theory, which states that at the 
Fermi energy scattering is completely elastic, and \cite{Nozieres}  
\be
t_\sigma(\omega=0^+)= 2\pi\varrho
\;{\cal T}_\sigma^{1CK}(\omega=0^+)=
2 \sin \delta_\sigma\;e^{i\delta_\sigma},
\label{eq:FL}
\ee
where we now allowed for a magnetic field pointing along the z-direction, 
and  $\delta_\sigma$ stands for the phase shifts of electrons with spin
$\sigma$ at the Fermi energy. Eq.~(\ref{eq:FL}) then yields
\bea 
&&\sigma_{\rm tot,\sigma}(\omega\to 0) = \frac {4\pi}{p_F^2}
\sin^2(\delta_\sigma)\;,
\\
&& \sigma_{\rm inel, \sigma}(\omega\to 0) = 0\;.
\eea
The maximum total scattering cross section is  reached in the unitary limit,
 $\delta_\sigma=\pi/2$. 

Let us now proceed and compute the various scattering cross section 
using 
Wilson's NRG~\cite{NRG_ref}.
In the previous section, we 
showed how the imaginary part of the single particle $T$-matrix is 
related to composite Fermion's spectral function, $\varrho_F(\omega)$. 
Within NRG, spectral functions of local operators are computed 
using their  Lehman representation. The imaginary part of the 
$T$-matrix, related to the total scattering rate 
has already been computed in this way by Costi to obtain the
magneto-resistivity of Kondo alloys.~\cite{Costi}  To evaluate the {\em inelastic}
scattering amplitude, however, one needs to go one step further and compute the
real part of the $T$-matrix as well through a Hilbert transformation, 
Eq.~(\ref{spectral}). In such a calculation it is essential to have high quality data.
The most challenging task is to obtain the correct $\sim\omega^2$ 
low energy behavior of the inelastic amplitude since we get this small
quantity as a difference of two quantities of the order of unity. Therefore it
is also crucial to get the normalization factor of $T$ correctly. 
In case of the single-channel Kondo problem
this can be obtained through the Fermi liquid relation, 
(\ref{eq:FL}): This relation 
connects
the normalization of 
$t(\omega)$ to the  phase shifts at the Fermi energy, which we 
extract 
from the NRG finite size spectrum very accurately\cite{Hofstetter}.

\begin{figure}
\includegraphics[width=8cm,clip]{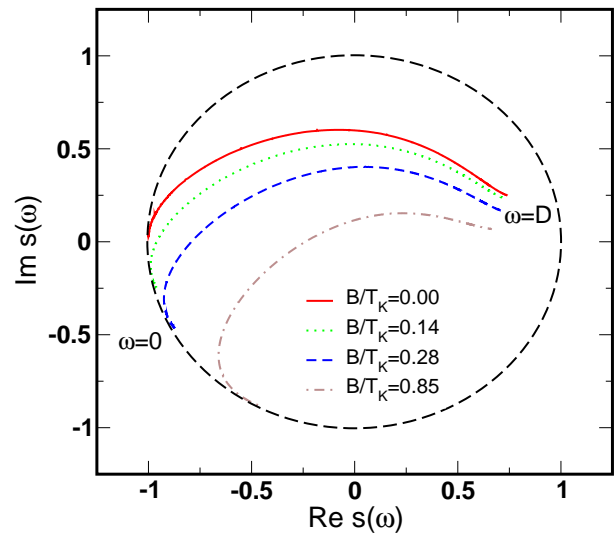}
\caption{
\label{fig:Sflow_1CK_inB}
Renormalization group flow of the eigenvalues of the single 
particle $S$-matrix for the single-channel Kondo model in presence of
a local magnetic field.}
\end{figure}
The renormalization group 
flow of the eigenvalues of the single 
particle $S$-matrix has already been shortly discussed in the introduction
 in the absence of magnetic fields (see Fig.\ref{fig:ineastic_elastic}). The
eigenvalues $s(\omega)$ of the $S$-matrix lie within the complex unit circle, 
and inelastic processes are 
allowed only when the eigenvalue $s(\omega)$ of the $S$-matrix 
satisfies $|s(\omega)|<1$.  Incoming particles of high enough energy 
($\omega\to\infty$) do not see the impurity, and  therefore $s(\omega)\to1$,
corresponding to the weak coupling fixed point with phase shift
$\delta\approx0$. As  expected for a Fermi liquids, at the Fermi energy 
($\omega\to0$) the $s(\omega)$ approaches the unit circle again,
$s(\omega=0)=-1$, corresponding to the strong coupling fixed point of
the Kondo model characterized by a phase shift $\delta=\pi/2$.

Application of a local magnetic field makes the 
flow more complicated (see Fig.\ref{fig:Sflow_1CK_inB}.): 
At low energy the system still behaves like a
Fermi liquid but the position of the point where the $s(\omega)$ approaches
the unit circle now varies with magnetic field. This is due to the 
 magnetic field dependence of the scattering phase shifts at
zero frequency.

\begin{figure}
\includegraphics[width=8cm,clip]{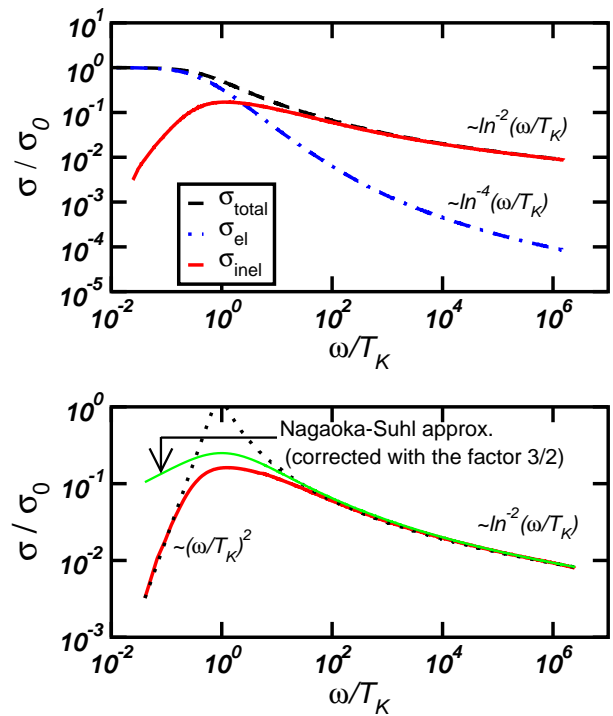}
\caption{
\label{fig:Inel_1CK}
  Upper panel: Inelastic, elastic, and
  total scattering 
  rates
  for the single-channel Kondo model
  in units of $\sigma_0 = 4\pi/p_F^2$
  at $T=0$ and 
  $B=0$,
  as a function of the 
  incoming electron's energy.   
  Only the
  electronic contribution ($\omega>0$) is plotted. 
  For $\omega
  \gg T_K$ the inelastic rate as well as the total scattering rate
  varies as
  ${\rm ln}^{-2}(T_K/\omega)$ while the elastic part
  decays as ${\rm ln}^{-4}(T_K/\omega)$.
  The lower panel shows the $\sim \omega^2$ and 
  ${\rm ln}^{-2}(T_K/\omega)$ regimes for $\omega \ll T_K$ and $\omega
  \gg T_K$, respectively. The Nagaoka-Suhl approximation, corrected with
  a factor $3/2$ (see the text) is also shown. 
}
\end{figure}

For intermediate
energy values of the incoming electron $|s(\omega)|<1$ and 
the inelastic scattering cross-section is non-zero. 
The total, elastic and inelastic scattering cross sections of an electron
scattered off a magnetic impurity are shown in Fig.\ref{fig:Inel_1CK}.
As expected, the
inelastic amplitude always vanishes at the Fermi level. In the lower panel we
show the inelastic scattering rate as compared to the Nagaoka-Suhl formula. 
The numerical results are consistent with the analytical 
expression (\ref{sigma_inel_asymp})  at large energies, while 
for energies much smaller than $T_K$ we recover the quadratically 
vanishing inelastic rate expected from Fermi liquid
theory.~\cite{NozieresII}
Note that
the Nagaoka-Suhl 
approximation systematically 
underestimates the inelastic scattering rate by a factor of 2/3
since it 
considers
any spin-diagonal process as elastic scattering.
At high energies, however, in leading order all the scatterings are inelastic
since even 
a
spin diagonal process breaks up the Kondo singlet
and leaves the system in an excited state, 
and therefore
it cannot be elastic.
Apart from this prefactor, 
the Nagaoka-Suhl result is perfect at high energies, however, it
starts to deviate strongly from the numerically exact curve at approximately
$10T_K$, and it completely fails below the Kondo temperature $T_K$.
At energies well above $T_K$ 
almost all the scattering is inelastic, i.e. the inelastic
amplitude varies as $\sim\ln^{-2}(\omega/T_K)$ while the elastic
part vanishes faster as $\sim\ln^{-4}(\omega/T_K)$, in agreement with the 
analytical results.

Even though the numerics recover the expected  asymptotics,
interesting features appear both in the low and high energy part
of the scattering properties. First, as shown in 
Fig.\ref{fig:sigma_summary}, 
the $\sigma_{\rm inel}\sim\omega^2$ regime appears only at
energies well below the Kondo temperature, and  we find that the inelastic
scattering rate is roughly linear between $0.05T_K<\omega<0.5T_K$.
Even though our calculation is done at  $T=0$ temperature, we expect that
$\sigma_{\rm inel}(T,\omega=0)$ behaves very similarly to
$\sigma_{\rm inel}(T=0,\omega)$. Our results
are thus 
consistent with the existing experimental 
data, 
explain the linear behavior observed in many 
experiments,~\cite{mohanty2003,Saminadayar2003,Saminadayar2005} 
and 
surprisingly
even quantitatively fit the 
finite temperature
experimental curves.~\cite{Saminadayar2005}
Of course, in reality a finite temperature calculation is needed
which has been performed in Ref.~\cite{Rosch}. 

Another remarkable feature is 
the broad pla\-teau in the inelastic
scattering cross section above the Kondo scale, where the 
energy-dependence of the inelastic scattering rate turns out to be 
 extremely weak. This weak energy-dependence  provides a natural explanation 
for the experimentally observed plateau of the dephasing rate 
in many experiments.~\cite{Mohanty-Webb,Birge2003}

\begin{figure}
\includegraphics[width=8cm,clip]{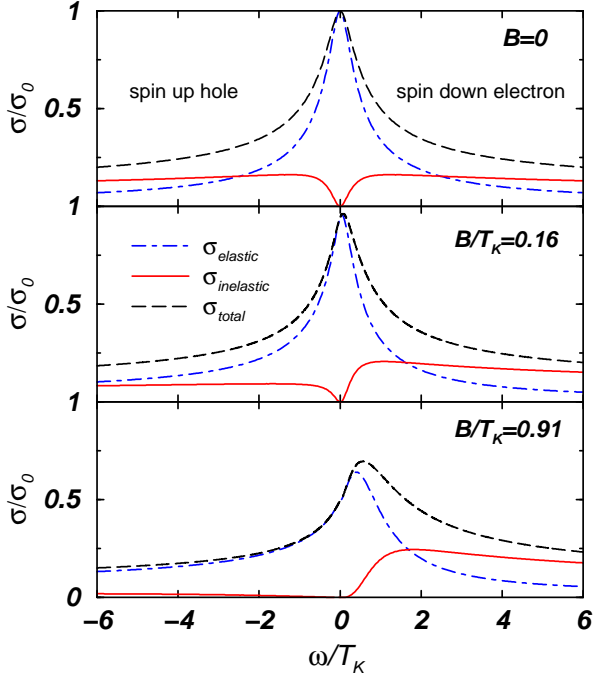}
\caption{
\label{fig:1CKinel_inB}
  Energy dependence of
  spin-dependent elastic and inelastic scattering rates 
  for the single-channel Kondo model
  in units of
  $\sigma_0 = 4\pi/p_F^2$, at $T=0$ and in the presence of a local magnetic
  field $B$. 
Note that the positive frequency side corresponds to spin down 
electrons,
while
the negative frequency side is for spin up holes. The impurity spin is
polarized upwards.}
\end{figure}
The inelastic scattering amplitude in
presence of a magnetic field is shown in 
Fig.\ref{fig:1CKinel_inB}.
Applying a  local magnetic field breaks the spin symmetry
of the scattering and changes the inelastic scattering properties
of spin up and down particles dramatically. 
Already a relatively small magnetic field $B\sim0.1T_K$ results in a 
very strong spin asymmetry of the inelastic
scattering. For $B\sim T_K$ the effect is even more dramatic: At this 
field the impurity is practically polarized and aligned with the
direction of the field and points upwards.
As a result, an incoming spin up particle cannot flip the local spin
in a first order process, and higher order processes are needed to generate 
inelastic 
scattering.  A spin down   electron or hole, on the other hand, 
can exchange its spin with the magnetic impurity,
resulting in the maximum of the inelastic rate at energy $\approx B$
and a very broad inelastic background for $\omega>B$. 

\section{Inelastic scattering in the two-channel Kondo model}
\label{sec:inel2CK}

In this section we shall present results for the two-channel Kondo model, 
the prototype of all non-Fermi liquid impurity 
models.~\cite{AndreiDestrii,AL} 
In the channel symmetric case there is
two 
types
of conduction electrons that try to screen the impurity
independently leading to the overscreening of the local
moment. This frustration of the screening processes manifests itself
in the formation of a strongly correlated state which cannot be
described in the framework of Fermi liquid theory. This unusual 
correlated state
manifests
in the nonzero residual entropy, the
logarithmic divergence of the impurity susceptibility and the power
law behavior of transport properties with fractional exponents. 

Since
the non-Fermi liquid behavior is a direct consequence of the
frustration of the screening processes, any infinitesimal 
asymmetry in the couplings
$\Delta\equiv(j_1-j_2)/(j_1+j_2)$ leads to
the appearance of another low temperature
energy scale $T^*\propto \Delta^2/T_K$ at which the system crosses over to 
a Fermi liquid: Electrons
being more strongly coupled to the impurity  form a usual Kondo
singlet with the impurity spin, while the other electron channel becomes
completely  decoupled from the spin.

In the 2-channel Kondo case, unfortunately, no Fermi-liquid
relations  similar to Eq.~(\ref{eq:FL}) are available. However, there is an
exact theorem by Maldacena and Ludwig, that allows us to get the right
normalization of the $T$-matrix. This theorem
states that, at the two-channel Kondo fixed point, the   
single-particle elements of the $S$-matrix 
vanish at the Fermi
energy, $s_{2CK}(\omega\to 0)= 0$,~\cite{Maldacena}
and as a consequence  
$$
t_{2CK} (\omega=0^+)= -i \;.
$$
This relation allows us to obtain the proper normalization of the 
numerically computed $T$-matrix even at the non-Fermi liquid fixed point. 
However, it also leads to the surprising result mentioned already in the 
introduction, that exactly half of the scattering is inelastic 
at the Fermi energy, while the other half of it is inelastic. 
This counter-intuitive result can be understood as follows: The 
identically zero single particle $S$-matrix indicates that
an incoming electron {\em cannot} be detected as one electron after 
the scattering event, and it ``decays'' into many electron-hole
pairs. To get such a ``decay'', the scattering process must have an
elastic 
component too
which interferes destructively with the not scattered
direct wave 
and results
in the absence of the outgoing single particle
amplitude in the $s$-channel.

\begin{figure}
\includegraphics[width=0.9\columnwidth,clip]{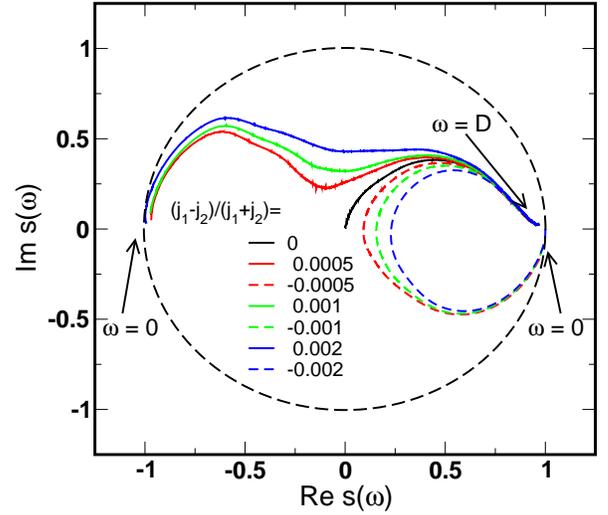}
\caption{\label{fig:2CK_Sflow}
Renormalization group flow of the eigenvalues of the single 
particle $S$-matrix for the two-channel Kondo model for 
various values of channel asymmetry.
The curves correspond to $j_1+j_2=0.12$.}
\end{figure}

The universal flow of the eigenvalue of the $S$-matrix was shown in 
Fig.~\ref{fig:Sflow}. 
In Fig.\ref{fig:2CK_Sflow} we show what happens  if we make the couplings in
the two channels slightly asymmetric.
For any small asymmetry the Fermi liquid 
behavior reappears: The $S$-matrix in the more strongly coupled channel 
flows first close to the two-channel Kondo fixed point with 
$s(\omega)\approx 0$, and then below the Fermi liquid scale 
$\omega < T^*$ it   suddenly crosses over 
to the strong coupling fixed point characterized with phase shifts
$\delta=\pi/2$ Similarly, 
$s(\omega)$ in
the other channel also approaches the 
two-channel Kondo fixed point, but then it becomes suddenly decoupled and
therefore $S$ flows to the $s(\omega)=1$ fixed point.

\begin{figure}
\includegraphics[width=8cm,clip]{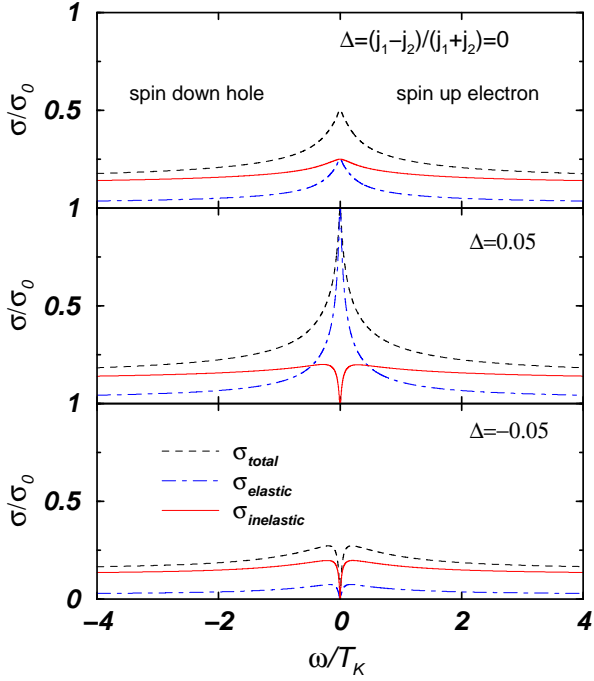}
\caption{\label{fig:2CK_inel}
Energy dependence of
  elastic and inelastic scattering rates 
  for the two-channel Kondo model
in channel ``1''   in units of
  $\sigma_0 = 4\pi/p_F^2$, at $T=0$ and 
  for different channel asymmetry, $\Delta\equiv(j_1-j_2)/(j_1+j_2)$.
The curves correspond to $j_1+j_2=0.3$.
}
\end{figure}
The inelastic scattering rates for the two-channel Kondo model are 
shown in Fig.~\ref{fig:2CK_inel} as a function of the energy of the incoming
particle. 
In the channel-symmetric case inelastic  processes are allowed  
even at $\omega=0$, which is a clear signature of the
non-Fermi liquid behavior. 
The non-Fermi liquid nature
is also reflected in the $\sim\sqrt{\omega}$ singularity of
the scattering cross sections at $\omega=0$. 
Note that this cusp is much 
less pronounced in the inelastic scattering rate.

For $\Delta>0$ the total scattering rate approaches the unitary limit
in channel ``1'' below the Fermi liquid scale $T^*$.
For $\Delta<0$, on the other hand, the total scattering rate goes to 0
in channel ``1'' below the Fermi liquid scale $T^*$.
In both cases, the inelastic scattering  freezes out, and 
$\sigma_{\rm inel}(\omega)$ shows a dip below $T^*$, 
and it ultimately scales to 0 as
$$
\sigma_{\rm inel}(\omega)\approx cst. \frac {\omega^2}{{T^*}^2}\;.
$$
Note that the inelastic scattering cross-section is 
very similar for $\Delta>0$  and  $\Delta<0$, while the total 
 scattering contributions 
differs 
dramatically 
in these two cases.

\section{Inelastic scattering in the Anderson model} 
\label{sec:inelAnderson}

As a final example, let us consider the Anderson model defined by the 
Hamiltonian, 
 \begin{eqnarray}
 H&=&\sum_{\vp\sigma}\epsilon(\vp) a^{\dagger}_{\vp
   \sigma}a^{\phantom{\dagger}}_{\vp\sigma}
+\epsilon_d\sum_{\sigma}d^{\dagger}_\sigma d^{\phantom{\dagger}}_\sigma
\nonumber \\ &+& 
Ud^{\dagger}_{\uparrow}d^{\phantom{\dagger}}_{\uparrow}d^{\dagger}_{\downarrow} d^{\phantom{\dagger}}_{\downarrow}
+ V \sum_{\sigma,\vp } \left( c^{\dagger}_{\vp \sigma}  d^{\phantom{\dagger}}_\sigma+\textrm{h.c.}\right)\;, 
 \end{eqnarray}
where now  $d_\sigma$ denotes the local d-level's annihilation operator, 
$U$ is the on-site Coulomb repulsion, and 
the conduction band and the local electronic level are hybridized by $V$. 

The $T$-matrix for the Anderson model can be related to the $d$-level's 
Green's function, as first discussed by Langreth.~\cite{langreth} 
The required relation can be trivially  established  
the the path integral formalism presented in Section~\ref{appen:tmatrixkondo}. 
The final result of this derivation, which is to some extent discussed in
Appendix~\ref{appen:tmatrixanderson} can be written as:
\bea
{\rm Im} \{ {\cal T}_{\sigma}(\omega)   \} &=&  \pi  {V^2}
\varrho_{d,\tau \sigma}(\omega)\;, 
\nonumber
\\
- \tau \ {\rm Re}\{ {\cal T}_{\sigma}(\omega)\} &=&  
  {V^2} \int d\omega'\; \frac{\varrho_{d,\tau \sigma}(\omega)}{\omega-\omega'}
\;, 
\label{eq:andersonT}
\eea
with $\varrho_{d,\sigma}(\omega)$ the spectral function of the $d$-Fermion's 
spectral function, and $\tau={\rm sgn}\; \xi$.

The ground state of the Anderson model is of a Fermi liquid. 
Therefore, the Fermi liquid relations (\ref{eq:FL}) can be used again to 
properly normalize the $T$-matrix. As we discussed in
Ref.~\onlinecite{zar_inel}, the Fermi liquid relations 
also
imply that at the Fermi energy the eigenvalue of the 
single particle $S$-matrix lies on the unit circle, and 
therefore the inelastic scattering rate vanishes.

\begin{figure}
\begin{center}
\includegraphics[width=0.9\columnwidth,clip]{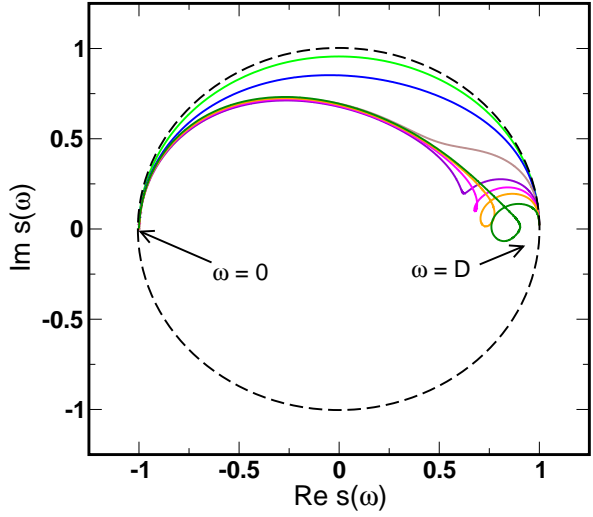}
\end{center}
\caption{
Flow of the eigenvalue  of the $S$-matrix for the Anderson model for
$\Delta=0.035$, $U=0.04$, $0.08$, $0.16$, $0.26$, $0.32$, $0.40$, $0.64$ and
$\varepsilon_d=-U/2$. 
}
\label{anderson_sflow}
\vspace{0.5cm}
\end{figure}
The flow of the eigenvalue $s(\omega)$ is shown in Fig.~\ref{anderson_sflow}. 
This flow diagram is very similar to that of the Kondo model at low energies,
however, a new interesting feature appears at 
$\omega\sim U$, where $s(\omega)$
displays a hook. This hook corresponds to largely 
inelastic scattering processes, which are associated with the 
{\em charge fluctuations} of the 
d-level. 
It is adequate to mention here that the low energy flow is not completely
identical to that shown in Fig.\ref{fig:Sflow} for the Kondo model. The 
reason is purely technical: For the Anderson model we were using
the self-energy trick invented by Bulla and coworkers\cite{ralf_selfenergy} 
to obtain higher quality results, and computed the
$\omega\sim T_K$ part of the $T$-matrix (especially its real part)
much more accurately.

\begin{figure}
\begin{center}
\includegraphics[width=0.9\columnwidth,clip]{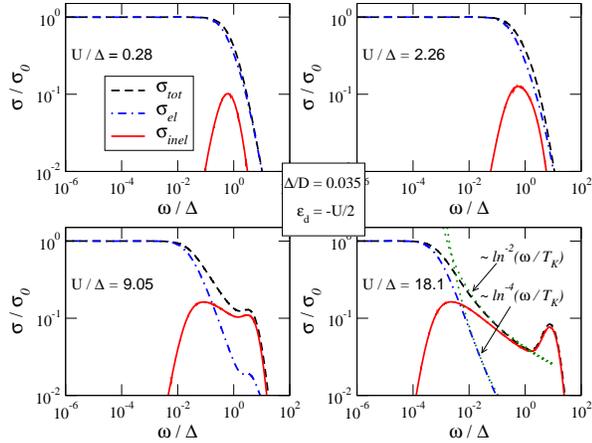}
\end{center}
\caption{
{Elastic, inelastic and total scattering cross section
for the symmetric Anderson model
$\varepsilon_d=-U/2$, for various ratios of 
$U/\Delta$. For moderate values of $U/\Delta$ the effects of U are
minor in the total and the elastic scattering rate, while
for large values of $U/\Delta$ the various scattering rates 
for $\omega\ll U$
follow very nicely the  behavior found for the Kondo model
}
\label{anderson_symm}}
\vspace{0.5cm}
\end{figure}

These features also appear in the various scattering rates, shown in 
Fig.~\ref{anderson_symm} for the symmetrical Anderson model with 
$\epsilon_d=-U/2$. There we show the inelastic scattering rate 
for various ratios of $U/\Delta$, $\Delta=\pi \varrho V^2$ being the 
width of the resonance. For moderate values of $U/\Delta$ the effects of U are
minor in the total and the elastic scattering rate, 
however, rather surprisingly, one can  see a clear maximum in the inelastic 
scattering rate  at energies $\omega\sim U$ even in this case. Increasing 
$U/\Delta$, the Fermi liquid regime and the charging regime separate, and 
two distinct peaks appear, now  even in the total and elastic scattering rates. 
For large values of $U/\Delta$ the various scattering rates 
follow very nicely the  behavior found for the Kondo model
at low energies, and for $T_K\ll\omega\ll U$ the elastic and   inelastic contributions 
scale as $\sim 1/\ln^4(\omega/T_K)$ and $\sim 1/\ln^2(\omega/T_K)$,
respectively. 
It is remarkable, that the Hubbard peak 
at $\omega\sim U$ is essentially entirely inelastic.

\begin{figure}
\begin{center}
\includegraphics[width=0.9\columnwidth,clip]{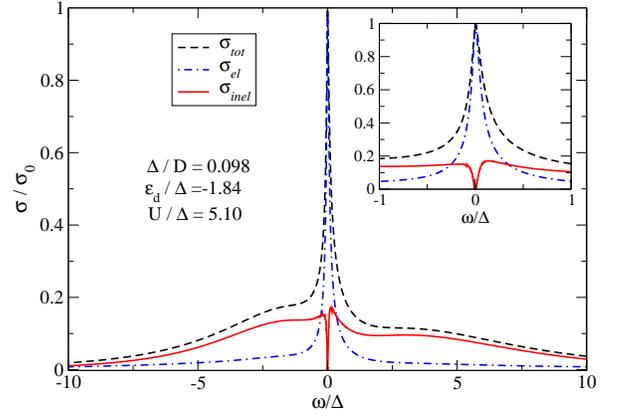}
\end{center}
\caption{
Elastic, inelastic and total scattering cross section
for the asymmetric Anderson model. The low energy
part of the curves is again very similar to the one 
obtained for the Kondo model.
}
\label{anderson_asym}
\vspace{0.5cm}
\end{figure}

Fig.~\ref{anderson_asym} shows the same behavior on a linear scale
for the asymmetrical Anderson model with moderate interaction strength. 
The low-energy part of the figure is again strikingly similar to the one 
obtained for the Kondo model. This is not very surprising, since 
the Kondo model is just the effective model of the Anderson 
 model in the limit of large $U/\Delta$ and $\omega \ll U$, where charge
 fluctuations occur only virtually. It is remarkable that 
the quasi-linear behavior of  $\sigma_{\rm inel}$ and the plateau are already 
present for these moderate values of $U/\Delta$.

\section{Conclusion}
\label{sec:conclusion}
In this paper, we discussed in detail the theory of inelastic scattering 
from quantum impurities at $T=0$ temperature, as formulated 
in Ref.~\onlinecite{zar_inel}, and  applied this formalism  
to various cases. 
We computed numerically the flow of the $S$-matrix eigenvalues $s(\omega)$ for three
prototypical examples of quantum impurity models, the Kondo model, the
two-channel Kondo model, and the Anderson model. As we discussed, inelastic
scattering appears, once $|s(\omega)|<1$, and the crucial difference
between Fermi liquid models and non-Fermi liquid models is that for non-Fermi
liquid models  $|s(\omega)|<1$ even at the Fermi energy, $\omega\to
0$, while for Fermi liquids $|s(\omega=0)|=1$. 

We also determined the inelastic scattering cross section, 
$\sigma_{\rm inel}(\omega)$, for all these models.
For the Kondo model and the  Anderson model in the Kondo regime, 
i.e., for large  interaction values, the low-energy 
part of $\sigma_{\rm inel}(\omega)$ 
has features essentially identical to those of the Kondo model:
Deep in the Fermi liquid regime one has 
$\sigma_{\rm inel}(\omega)\sim (\omega/T_K)^2$, while for $0.05 T_K<\omega<0.5
T_K$ a quasi-linear regime appears, above which  $\sigma_{\rm inel}$ exhibits a
plateau with  $\sigma_{\rm inel}(\omega)\approx cst$ over a wide frequency
range. These features are quite robust, and survive even in the case of
electron-hole symmetry breaking.

We also find that at large frequencies the scattering becomes asymptotically 
inelastic, and the inelastic scattering rate scales as 
$$ 
\frac 1 {\tau_{\rm inel}}  \approx n_{\rm imp} 
\frac {  \pi  S(S+1) } {2\;\varrho \;\ln^2(\omega/T_K)}\;.
$$ 
while the elastic scattering rate falls off much more rapidly as
$$ 
\frac 1 {\tau_{\rm el}}  \approx n_{\rm imp} 
\frac {  \pi^3  S(S+1) } {8\;\varrho \;\ln^4(\omega/T_K)}\;.
$$
This result implies that --contrary to common wisdom-- even 
spin-diagonal scattering is inelastic at high energies.~\cite{Garst}

In addition to these remarkable low-energy features, the  Anderson Hamiltonian 
exhibits another, very interesting inelastic scattering peak at $\omega\sim U$
that corresponds to {\em charge excitations}. Rather surprisingly, this 
 peak is present even in the weak coupling regime, where no Hubbard peak can
 be seen in the total scattering cross section. In the Kondo regime, 
on the other hand, 
this peak is essentially identical 
to 
the 
 Hubbard peak that appears in the total scattering 
cross-section, and which corresponds to almost completely 
inelastic scattering.

In the two-channel Kondo model, the prototype of all non-Fermi liquid models,
inelastic scattering remains finite even if $\omega\to 0$, and is exactly half
of the total scattering rate. However, the tiniest channel symmetry breaking 
destroys this non-Fermi liquid state, and generates a new 
Fermi liquid scale, $T^*$, below which inelastic scattering freezes out, and
the scattering becomes totally elastic.

The inelastic scattering rates computed here for the Kondo and Anderson models 
and their finite temperature versions computed in Ref.~\onlinecite{Rosch}
 are in 
{\em quantitative}  agreement with recent experimental studies
on magnetically doped mesoscopic wires  excepting the 
limit of
very small temperature,
where a small residual inelastic scattering rate seems to be 
present.~\cite{Saminadayar2005,Saminadayar2006,Birge2006}   
The origin of this small residual inelastic scattering rate is 
not clear yet, it might be due to some structural defects caused 
by the implantation process,
or just some magnetic ions located at the interface of the wire. 
The agreement is even more surprising, since in
reality,  magnetic impurities are {\em not} of spin $S=1/2$ character, but 
have a rather complicated $d$-level 
structure\cite{Zawareview}. They thus usually have a 
large spin associated with them (typically $S=2$ or $S=5/2$ for 
for Fe, Cr, or Mn)  subject to crystal fields, that does not couple through
a simple exchange interaction to the conduction electrons. In reality
scattering thus takes place in some $d$-electron channels. For 
$S=5/2$, e.g., the Fermi liquid state forms due to screening in 
five $d$-channels. Unfortunately, these realistic impurity models are 
out of reach for NRG computations. 

In case of $d$-impurities, scattering 
cross-sections become also larger due to the many angular momentum channels 
that are open to scattering. Assuming spherical symmetry, e.g., 
the 
angle averaged
total and elastic scattering cross sections become  
\bea
\sigma_{\rm tot} &=& \frac {2\pi}{p_F^2}\sum_{L} \; {\rm Im}\{t_L(\omega)\} 
\\
\sigma_{\rm el} &=& \frac {\pi}{p_F^2}\sum_{L} |t_L(\omega)|^2\;,
\eea
i.e., the total, elastic and inelastic scattering cross-sections are 
about {\em five times larger} for $d$-wave scattering than for 
$s$-wave scattering considered in the usual Kondo problem. This must also 
be taken into account when computing the amplitude of the 
 observed Kondo anomaly or that of the inelastic scattering rate. Finally, 
band structure effects may also play an important role in real materials, 
where the Fermi surface is not spherical, and the Fermi velocity depends 
on the direction of incidence\cite{Achim2}.

\acknowledgments
We are indebted to  L. Saminadayar, C. B\"auerle, J.J. Lin, and A. Rosch 
for valuable discussions. 
This research has been supported by Hungarian grants Nos.
NF061726, D048665, T046303 and  T048782,
by the DFG center for functional nanostructures,
(CFN), German Grant No. DFG SFB 608, and 
by the Alexander von Humboldt Foundation.
G.Z. acknowledges the
hospitality of the Center of Advanced Studies, Oslo.
L.B. acknowledges the financial support of the Bolyai 
Foundation.

\appendix

\section{Field-Theoretic derivation of the T-matrix 
for the Anderson model}

\label{appen:tmatrixanderson}
Here
we derive the T-matrix for the Anderson model following the lines of 
Sec.~\ref{appen:tmatrixkondo}. 
We first introduce the generating functional for the Green's functions

 \begin{eqnarray}
 Z(\eta,\overline{\eta})= \int D[\overline{a}_\sigma a_\sigma] D[\overline{d}_\sigma d_\sigma]e^{-iS}e^{i \overline{\eta} a + i \eta \overline{a}},
 \end{eqnarray}
 where the source terms are defined as in 
Sec.~\ref{appen:tmatrixkondo}. The 
action $S=S_{e}+S_{d}+S_{hyb}+S_{int}$ consists of four distinct parts, 
with
$S_e$ 
defined by Eq.\eqref{eq:Se}
is identical to the one given in Sec.\ref{appen:tmatrixkondo} and 
the remaining parts 
given by
 \begin{eqnarray}
 S_d&=&- \sum_\sigma \int d t d t' \overline{d}_\sigma (t) \left [G^0_d\right]_\sigma (t-t') d_\sigma (t') \; , \nonumber \\  S_{hyb} &=& V \sum_\sigma \int d t  \left( \overline{a}_\sigma (t,0) d_\sigma (t)+\textrm{c.c}\right) \; , \nonumber \\ S_{int}&=& U \int d t \overline{d}_{\uparrow}(t) \overline{d}_\downarrow (t) d_\downarrow (t) d_\uparrow (t)\; ,
 \end{eqnarray}
 respectively. 
Shifting the integration variables 
generates the
terms
\begin{widetext}
\begin{eqnarray}
&-& \overline{\eta} \cdot a-\eta \cdot \overline{a}  \to  \sum_{{\bf {p}},\sigma} \int d t d t' \overline{\eta}_{{\bf{p}},\sigma} G^0_{{\bf{p}}\sigma} (t-t') \eta_{{\bf{p}},\sigma} -V \sum_{{\bf {p}},\sigma} \int d t d t' G^0_{{\bf{p}}\sigma} (t-t') \left ( \overline{\eta}_{{\bf {p}},\sigma} (t') d_\sigma (t)+ \textrm{c.c.} \right)  \; , \nonumber \\ 
\end{eqnarray}
and which,
 after functional differentiation with respect to
$\eta_{{\bf{p}},\sigma}$, 
give rise to the identity
\begin{eqnarray}
&&G_{{\bf{p}}\sigma,{\bf{p'}}\sigma'}(t-t')= \delta_{{\bf{p}},{\bf{p'}}} \delta_{\sigma,\sigma'} G^0_{{\bf{p}}\sigma}(t-t') -i V^2 \delta_{\sigma,\sigma'}\int d \tilde{t} d \tilde{t}' G^0_{{\bf{p}}\sigma}(t-\tilde{t}) \langle d_{\sigma} (\tilde{t}) \overline{d}_{\sigma '} (\tilde{t}')\rangle G^0_{{\bf{p'}}\sigma' }(\tilde{t}'-t'). \nonumber \\
\end{eqnarray}
The Fourier transform of this equation yields
Eq.\eqref{eq:andersonT}.
\end{widetext}



\begin{thebibliography}{99}
\bibitem{NozieresII} In Ref.~\cite{Nozieres} Nozi\`eres considers stable 
quasiparticles rather than electrons, which do not decay at $T=0$.
\end{thebibliography}
\end{document}